\documentclass[smallextended,referee,envcountsect]{svjour3}

\smartqed

\usepackage{amsfonts}
\usepackage{pdfsync}
\usepackage{placeins}
\usepackage{graphicx}
\usepackage{amsmath}
\usepackage{setspace}
\usepackage[english]{babel}
\usepackage[left=3cm,right=3cm,top=2.5cm,bottom=3cm]{geometry}
\usepackage{amsmath,amssymb,bbm,color,graphics,verbatim,version}
\usepackage{mathrsfs}

\newcommand{\cA}{\mathcal{A}}

\newcommand{\E}{\mathbb{E}}

\newcommand{\R}{\mathbb{R}}
\newcommand{\p}{\mathbb{P}}

\newcommand\h{H_{q}} \newcommand\hp{H_q^\prime}
\newcommand\w{W_{q}} \newcommand\wprime{W_q^\prime}
\newcommand\gqw{G_{q,w}}

\def\WT{\widetilde}

\begin{document}


\title{On the Optimal Dividend Problem for Insurance Risk Models with Surplus-Dependent Premiums}

\author{Ewa Marciniak   \and  Zbigniew Palmowski }

\institute{Ewa Marciniak \at
             AGH University of Science and Technology \\
              Krak\'ow, Poland
           \and
           Zbigniew Palmowski,  Corresponding author  \at
               University of Wroc{\l}aw \\
              Wroc{\l}aw, Poland\\
              \email{zbigniew.palmowski@gmail.com}
}

\date{Received: date / Accepted: date}

\maketitle

\begin{abstract}
This paper concerns an optimal dividend distribution problem for an
insurance company
with surplus-dependent premium.
 In the absence of dividend payments, such a risk process is a particular case of so-called piecewise deterministic Markov processes.
The control mechanism chooses the size of dividend payments. The objective consists in maximazing the sum of
the expected cumulative discounted dividend payments received until the
time of ruin and a penalty payment at the time of ruin, which
is an increasing function of the size of the shortfall at ruin.
A complete solution is presented to the corresponding stochastic
control problem.
We identify the associated Hamilton-Jacobi-Bellman equation and find
 necessary and sufficient conditions for optimality
of a single dividend-band strategy, in terms of  particular Gerber-Shiu functions.
A number of concrete examples are analyzed.
\end{abstract}

\keywords{  optimal strategy $\star$ PDMP $\star$ barrier strategy $\star$ integro-differential HJB equation $\star$ Gerber-Shiu function $\star$ stochastic control}
\subclass{60G51, 60G50, 60K25, 93E20}

\section{Introduction}
In classical collective risk theory the surplus
of an insurance company is described by
the Cram\'er-Lundberg model.
Under the assumption that the premium income per unit time  is larger than the average
amount claimed, the surplus in the Cram\'er-Lundberg model has positive
first moment and has therefore the unrealistic property that it converges to infinity
with probability one.
In answer to this objection De Finetti \cite{finetti} introduced the dividend
barrier model, in which all surpluses above a given level are transferred to a
beneficiary, and raised
  the question of optimizing this barrier.
In the mathematical finance and actuarial literature, there is a good deal
of work
  being done on dividend barrier models and the problem of finding an optimal policy
of paying dividends. Gerber and Shiu \cite{GerberEllias} and Jeanblanc and Shiryaev \cite{JeanShir}
consider the optimal dividend problem in a Brownian setting. Irb\"{a}ck \cite{Irback} and Zhou
\cite{Zhou} study constant barriers. Asmussen et al.
\cite{AHT} investigate excess-of-loss reinsurance and dividend distribution policies in
a diffusion setting. Azcue and Muler \cite{AM}
take a viscosity approach to investigate
optimal reinsurance and dividend policies in the Cram\'er-Lundberg model
using a Hamilton-Jacobi-Bellman (HJB) system of equations.
Avram et al. \cite{APP,APP2}, Kyprianou and Palmowski \cite{KPal}, Loeffen \cite{Loeffen1,Loeffen2},  Loeffen and Renaud \cite{LR}
and many other authors
analyze the L\'evy risk processes set-up from the probabilistic point of view.

In this paper, we shall approach the dividend problem for a
reserve-dependent risk process using
  the  theory of piecewise deterministic Markov processes (PDMP). We also take into account
  the ``severity'' of ruin and therefore we consider
  the so-called
Gerber-Shiu penalty function
  (see e.g. Schmidli \cite{Schmbook} or Avram et al. \cite{APP2} and references therein).
For this set-up, without transaction costs, we find
  the corresponding HJB system. We analyze the barrier strategy for which all surpluses above a given level are transferred to dividends. In particular, we find necessary and sufficient conditions for the barrier strategy to be optimal.

We believe that
  PDMP models can  better describe the situation of an insurance company, since for example
  they can
invest the surplus into a bond with a fixed interest rate. Such a situation is described by
  a PDMP model with a linear premium (see~\cite{Seg1942}).

The paper is organized as follows.
In Section \ref{sec:model}, we introduce the basic
  notation and we describe the model we deal with. Section \ref{sec:pre} is dedicated to the
  related one-sided and two-sided problems.
In Section \ref{sec:main}, we present the
  Verification Theorem, necessary and sufficient conditions for the barrier strategy to be
  optimal.
In Section \ref{sec:proof}, we give all the proofs.
Section \ref{sec:examples} and \ref{sec:remarks} are devoted to some examples and concluding remarks.

\section{The Model}\label{sec:model}

In this paper, we assume that the surplus $R$ of an insurance company (without payment of
  dividends) with an initial capital $x$ is described by the following differential equation:
\begin{equation}\label{R}
R_t=x+\int_0^t p(R_s)\,ds-\sum_{k=1}^{N(t)}C_k,
\end{equation}
where $p$ is a given deterministic positive premium function, $\left\{C_k\right\}_{i=1}^\infty$ is a sequence of i.i.d. positive random variables with d.f. $F$ representing the claims, and $N$
is an independent Poisson process with intensity $\lambda$ modeling the times at which the
claims occur.

We assume that $R_t\to\infty$ a.s., $\E C<\infty$ for a generic claim $C$, and the premium rate $p$ is monotone, absolutely continuous and
satisfies the following ``speed condition'':
\begin{equation}\label{speed}
\int_0^\infty e^{-qt}p(r_t^x)\,dt\leq A x+B
\end{equation}
for some constants $A, B\geq 0$ and a function $r^x$ satisfying the equation
\begin{equation}\label{rtx}
r_t^x=x+\int_0^t p(r_s^x)\,ds.\end{equation}
Note that $r^x$ describes a deterministic trajectory of $R$ along which no claims appear.

\begin{remark}
\rm Constant and linear premium functions  satisfy the above assumptions.
For a constant premium function we obtain the classical Cram\'er-Lundberg model.
\end{remark}

To approach the dividend problem, we consider the regulated risk process satisfying the following stochastic differential equation:
\begin{equation}\label{Nasz}
X^\pi_t=x+\int^t_0p(X_s^\pi)\,ds-\sum_{k=1}^{N(t)}C_k-L^\pi_t,
\end{equation}
where $\pi$ denotes a strategy chosen from the class $\Pi$
of all ``admissible'' dividend  controls resulting  in the cumulative
amounts of dividends $L^{\pi}_t$ paid up to time $t$.
Note that ruin may be either exogeneous or endogeneous (i.e.,
caused by a claim or by a dividend payment). A dividend strategy
is admissible, if ruin is always exogeneous or, more precisely, an
admissible dividend strategy $L^\pi=\{L^\pi_t, t\in\R_+\}$ is a right-continuous stochastic process, adapted  to the natural filtration
of the risk process $R$  that satisfies the usual conditions, and such that, at any time
preceding the epoch of ruin, the dividend payment
is smaller than the size of the
available reserves ($L^\pi_t-L^\pi_{t-}<X_{t-}^\pi$).

The object of interest is the discounted cumulative dividend paid up to the ruin time:
$$
D(\pi):=\int_0^{T^\pi}e^{-qt}\,dL^{\pi}_t,
$$
where $T^\pi:=\inf\{t \geq 0: X^{\pi}_t<0\}$
is the ruin time and $q\ge0$ is a given discount rate. Note that unless it is necessary we will write $T$ instead of $T^\pi$ to simplify the notation.
The objective is to maximize $\E_x D(\pi)$, where $\E_x$ is the expectation with respect to $\p_x(\cdot)=\p(\cdot|X^\pi_0=x)$.
We will use the notation $\p_0=\p$ and $\E_0=\E$.

To take into account the ``severity'' of ruin, we also consider the so-called
Gerber-Shiu penalty function
\begin{equation*}\label{gerbershiu}
\E_x[e^{-qT}w(X^\pi_T)\mathbb{I}_{\{ T<\infty\}}]
\end{equation*}
for some general non-positive penalty function $w$ satisfing the integrability condition
$$\sup_{y\geq 0}\E\left[-w(y-C_1)                                                                                                                                                                                                                                                                                                                                                                                                                                                                                                                                                              |C_1>y\right]<\infty.$$
Note that for $q=0$ and $w=-1$ we derive the ruin probability.

The dividend problem consists in finding the so-called value function $v$ given by
\begin{equation}\label{cost}
v(x):=\sup_{\pi\in \Pi}v_\pi(x),
\end{equation}
where \begin{equation}\label{cost2}v_\pi(x):=\E_x \bigg[\int_0^{T}e^{-qt}\,dL^{\pi}_t+e^{-qT}w(X^\pi_T)\mathbb{I}_{\{ T<\infty\}}\bigg]
\end{equation}
and the optimal strategy $\pi_*\in\Pi$ such that
\begin{equation*}
v(x)=v_{\pi_*}(x)\qquad\text{for all }x\geq 0.
\end{equation*}
\section{Preliminaries}\label{sec:pre}

For the solution of the dividend problem, two functions, $W_q$ and $\gqw$, are crucial. They are related to two-sided and one-sided exit problems for $R$:
\begin{gather}\label{exit1}
\E_x\left[e^{-q\tau_a^+}\mathbb{I}_{\{\tau_a^+<\tau^-_0\}}\right]=\frac{W_q(x)}{W_q(a)},
\\
\label{exit2}
\gqw (x):=\E_x\left[e^{-q\tau_0^-}w(R_{\tau^-_0})\mathbb{I}_{\{\tau_0^-<\infty\}}\right],
\end{gather}
where $x\in]0,a[$, $\tau_a^+:=\inf\{t\geq 0: R_t\geq a\}$ and $\tau_0^-:=\inf\{t\geq 0: R_t<0\}$.
From now on we will assume the existence of the function $W_q$, which follows for example from the existence of the following limit:
$W_q(x)=\lim_{y\to\infty}\E_x [e^{-q \tau_y^+}\mathbb{I}_{\{ \tau_y^+<\tau_0^-\}}]/\E [ e^{-q \tau_y^+}\mathbb{I}_ {\{\tau_y^+<\tau_0^-\}}]$. Indeed, using the strong Markov property of $R$ that has only negative jumps,
we derive
$$W_q(x)=\lim_{y\to\infty}\frac{\E_x \left[e^{-q \tau_a^+}\mathbb{I}_{\{ \tau_a^+<\tau_0^-\}}\right]\E_a \left[e^{-q \tau_a^+}\mathbb{I}_{\{ \tau_y^+<\tau_0^-\}}\right]}{\E\left[e^{-q \tau_y^+}\mathbb{I}_{\{ \tau_y^+<\tau_0^-\}}\right]} =\E_x \left[e^{-q \tau_a^+}\mathbb{I}_{\{ \tau_a^+<\tau_0^-\}} \right]W_q(a),
$$
which gives the required identity (\ref{exit1}).

For the properties of the function $\gqw$ we refer the reader to \cite{Corina}, where numerous examples are studied.

\section{Main Results}\label{sec:main}
To prove the optimality of a particular strategy $\pi$ among all admissible strategies $\Pi$
for the dividend problem (\ref{cost}), we  consider the following Hamilton-Jacobi-Bellman (HJB) system:
\begin{equation}\label{HJB}
 \max\left\{ \cA m(x)-qm(x),1-m^\prime(x)\right\}\leq0\quad \text{for all }x>0,\qquad m(x)=w(x)\quad\text{for all }x<0,
\end{equation}
where $\cA$ is the full generator of $R$,
\begin{equation*}
\cA m(x)=p(x)m'(x)+\lambda\int_0^\infty(m(x-y)-m(x))\, dF(y),
\end{equation*}
acting on absolutely continuous functions $m$ such that
\begin{equation*}\label{domain}
\E\bigg[\sum_{\sigma_i\leq t}|m(R_{\sigma_i})-m(R_{\sigma_i-})|\bigg]<\infty
\qquad\text{for any}\ t\geq 0,
\end{equation*}
where $\{\sigma_i\}_{i\in \mathbb{N}\cup\{0\}}$ denotes the times at which the claims occur (see Davis \cite{Davis} and Rolski et al. \cite{Rolski}). In this case $m'$ denotes the  density of~$m$.
Note that any function, which is absolutely continuous and ultimately dominated by an affine function, is in the domain of the full generator $\cA$, as a consequence
of the assumption that $\E C_1<\infty$.
Recall that, for any function $m$ from the domain of  $\cA$, the process
$$\bigg\{e^{-qt}m(R_t) -\int_0^te^{-qs}\left(\cA-q\right) m(R_s)\,ds, t\geq 0\bigg\}$$
is a martingale.

\begin{theorem}[Verification Theorem]\label{verthm}
Let $\pi$ be an admissible dividend strategy such that $v_\pi$ is absolutely continuous and ultimately dominated by some affine function.
If \eqref{HJB} holds for $v_\pi$ then $v_\pi(x)=v(x)$ for all $x\geq 0$.
\end{theorem}

The proof of all theorems given here will be given in Section \ref{sec:proof}.

\begin{lemma}\label{diff}
Assume that the distribution function (d.f.) $F$ of the claim size is absolutely continuous. Then the functions $W_q$ and $G_{q,w}$ are continuously differentiable for all $x\geq 0$.
\end{lemma}

From now on we assume that the claim size distribution is absolutely continuous with a density~$f$.

We will focus on the so-called barrier
policy $\pi_a$ transferring all surpluses above a given level $a$ to dividends.

\begin{theorem}\label{barthm}
We have
\begin{equation}\label{va}
v_{a}(x) := v_{\pi_{a}}(x) =
\begin{cases} \frac{\w (x)}{\wprime (a)}(1-\gqw^\prime(a))+ \gqw(x),
& x \leq a,\\
x - a + v_a(a), & x >  a.
\end{cases}
\end{equation}
Moreover $v_{a}$ is continuously differentiable for all $x\geq 0$.
\end{theorem}

Let
$$ \hp(y):= \frac{1-\gqw^\prime(y)}{\wprime (y)}.$$
Define now a candidate for the optimal dividend barrier by
\begin{equation}
 a^*:=\sup\left\{a\geq0:\hp(a)\geq\hp(x)\text{ for all }x\geq0\right\},
\end{equation}
where $\hp(0)=\lim_{x\downarrow 0}\hp(x)$.

Finally, using the above two theorems we can give necessary and sufficient conditions
for the barrier strategy to be optimal.

\begin{theorem}\label{ver2}
The value function under the barrier strategy $\pi_{a^*}$ is in the domain of the full generator $\cA$.
The barrier policy $\pi_{a^*}$ is optimal and $v_{a^*}(x)=v(x)$ for all $x\geq 0$ if and only if
\begin{equation}\label{inq2}
 (\cA -q)v_{a^*}(x)\leq 0\qquad \text{for all }x>a^*.
\end{equation}
 \end{theorem}

\begin{theorem}\label{suffcond}
Suppose that
\begin{equation}\label{as1}
 \hp(a)\geq\hp(b)\qquad\text{for all } a^*\leq a\leq b.
\end{equation}
Then the barrier strategy at $a^*$ is optimal, that is, $v(x)=v_{a^*}(x)$ for all $x\geq 0$.
\end{theorem}

\begin{theorem}\label{convex}
Suppose that $f$ is convex and  $p$ is concave.
Then the barrier strategy at $a^*$ is optimal, that is, $v(x)=v_{a^*}(x)$ for all $x\geq 0$.
\end{theorem}

\begin{theorem}\label{suffcond2}
Consider the problem without the penalty function $(w\equiv 0)$.
Suppose that $f$ is  decreasing and
$$
p'(x)\leq q+\lambda,\qquad x\geq a^*,
$$
where $p'$  is the density of the premium rate $p$. Then the barrier strategy at $a^*$ is optimal, that is, $v(x)=v_{a^*}(x)$ for all $x\geq 0$.
\end{theorem}

\section{Examples}\label{sec:examples}

In this section, we will assume that the premium function $p$ is differentiable and the generic claim size has a density $f$
with a rational Laplace
transform. That is, there exists $m\in \mathbb{N}$ and constants $\left\{\beta_i\right\}_{i=0}^{m-1}$ such that the density $f$ satisfies the following LODE:
\begin{equation*}
  \label{eq:x}\mathcal{L}\left(\frac{d}{dy}\right)f(y)=0
  \end{equation*} with initial conditions
$\smash{f^{(k)}(0)=0} \: (k=0,\ldots, m-2)$, where
\begin{equation*}
  \mathcal{L}(x) =x^m + \beta_{m-1}x^{m-1} +\cdots+\beta_0.
\end{equation*}
Note that by Theorem \ref{convex} if we take $p$ concave then the barrier strategy is optimal for an exponential claim size (in this case $\mathcal{L}(x)=x+\mu$).
From Lemma \ref{diff} and its proof it follows that if the claim size distribution is absolutely continuous then $W_q$, $\gqw$ and $v_{a^*}$
are differentiable and satisfy
\begin{equation}\label{eq:h}
\cA W_q(x)=qW_q (x)\quad \text{for } x\geq 0,\qquad  W_q (x)=0\quad \text{for } x<0,
\end{equation}
and
\begin{equation}\label{eq:G}
\cA \gqw (x)=q\gqw (x) \quad \text{for }x\geq 0,\qquad  \gqw (x)=w(x)\quad \text{for } x<0.
\end{equation}
Our goal will be to find the value function $v$ for a few examples of  premium functions.
The Gerber-Shiu function $\gqw$ was
determined in Albrecher et al. \cite{Corina}.
One can prove that if $\gqw$ is differentiable then in fact $\gqw\in\mathcal{C}^{m+1}$ (see \cite{Rolski}).
The same holds for~$W_q$.
Albrecher et al.~\cite{Corina} proved that $\gqw$ satisfies
the following LODE with variable coefficients of order $m+1$:
\begin{equation}
  \label{eq:ode}
  \textbf{T} \gqw(x) = u(x)
\end{equation}
with the differential operator
\begin{equation*}\label{opT}
\textbf{T} := \mathcal{L}\left(\frac{d}{dx}\right)\left(q-p(x)\frac{d}{dx}+\lambda\right) -\lambda \beta_0
\end{equation*}
and the right-hand side $$
u(x):=\lambda
\mathcal{L}\left(\frac{d}{dx}\right)  \omega(x),$$
where $\omega(x):=\int_x^\infty w(x-z)\,dF(z)$.
In general, the  main idea of solving the above equation is to find stable solutions $s_k$ of the fundamental system for (\ref{eq:ode})
(that is, those vanishing at infinity) and
then use the representation
\begin{equation*}
\gqw (x) = \gamma_1 s_1(x)+\cdots+\gamma_m s_m(x) +Gu(x),
\end{equation*}
where $G$ is the Green operator
and the constants $\gamma_i$ can be computed from the initial conditions.
Moreover, the form of the Green operator is
found in \cite[Thm. 3.4]{Corina}.

If the claim size has exponential distribution with intensity $\mu$ then we can prove that $\gqw$ solves the following ODE:
\begin{equation*}
  \left(\frac{d^2}{du^2}+\left(\mu +\frac{p^\prime(x)}{p(x)}-\frac{\lambda
        +q}{p(x)}\right)\frac{d}{du} - \frac{q \mu}{p(x)} \right) \gqw (x) = u(x),
\end{equation*}
with $u(x)= -\frac{\lambda}{p(x)} (\frac{d}{du} +\mu) \, \omega(x)$.
This allows one to find $\gqw$ explicitly.

Moreover, note that (\ref{eq:h}) is a Gerber-Shiu function with zero penalty function.
In contrast to the  one from (\ref{exit1}) we now have $\lim_{x\to\infty} W_q (x)=+\infty$.
This means that the optimal value function under mild conditions
is a linear combination of two Gerber-Shiu functions: an unstable one that vanishes on the negative half-line and tends to infinity at infinity
(corresponding to dividend payment, $W_q$ in our notation), and
a stable one, vanishing at infinity (corresponding to the penalty payment, $\gqw$ in our notation).
From \cite{Corina} we know that $W_q$ equals  the unstable solution of the fundamental system for (\ref{eq:ode}).
One can prove that there exists a unique unstable solution (see \cite{Corina} for details).
In the rest of this section we will assume that the claim size has exponential distribution with  intensity~$\mu$.

\subsection{Linear Premium}
We take here $p(x)=c+\epsilon x$.
By Theorem \ref{convex} the barrier strategy at $a^*$ is optimal.
In this case
$$
s_1(x) = U\big(\tfrac{q}{\epsilon}+1, \,
    \tfrac{\lambda+q}{\epsilon} +1, \, \mu x + \tfrac{\mu
      c}{\epsilon} \big)\, ( \epsilon x +
    c)^{(\lambda+q)/\epsilon}  \exp(-\mu x)
$$
and
\begin{multline*}
   Gu(x)=
  \tfrac{\Gamma(q/\epsilon +
    1)}{\Gamma((q + \lambda)/(1+\epsilon))}
  \tfrac{1}{\epsilon}
  \left(\tfrac{\mu}{\epsilon}\right)^{(\lambda+q)/\epsilon}
  (\epsilon x +c)^{(\lambda+q)/\epsilon} \exp(-\mu x
  -\tfrac{\mu c}{\epsilon})\times \\
  \quad \Big(-U(x) \int_0^x M(v)u(v)\,dv - M(x) \int_x^{\infty} U(v)u(v)\,dv +
  \tfrac{M(0)}{U(0)} \, U(x) \int_0^{\infty} U(v)u(v)\,dv\Big)\, ,
\end{multline*}
where $U(u)$ and $M(u)$ are  Kummer functions.
This gives
$$\gqw (x)=s_1(x) +Gu(x)$$
for $u(x)=
-\frac{\lambda}{p(x)} (\frac{d}{du} +\mu) \, \omega(x)$.
Moreover,
\begin{align*}
W_q(x)&=C_1 M\big(\tfrac{q}{\epsilon}+1, \,
    \tfrac{\lambda+q}{\epsilon} +1, \, \mu x + \tfrac{\mu
      c}{\epsilon} \big)\, ( \epsilon x +
    c)^{(\lambda+q)/\epsilon} \, \exp(-\mu x)\\
    &\quad+C_2 U\big(\tfrac{q}{\epsilon}+1, \,
    \tfrac{\lambda+q}{\epsilon} +1, \, \mu x + \tfrac{\mu
      c}{\epsilon} \big)\, ( \epsilon x +
    c)^{(\lambda+q)/\epsilon} \, \exp(-\mu x),
\end{align*}
with $C_1$ and $C_2$ determined by the boundary conditions $W_q(0)=1$ and $W_q'(0)=(\lambda+q)/c$.

Hence we can find the optimal barrier $a^*$ by solving $\h''(a^*)=0$. In the case of absence of the penalty function, that is, when $w(x)=0$,
we can perform some numerical analysis of the values of~$a^*$. In  Tables 1, 2 and 3 we present some values of $a^*$ for different parameters.
\medskip
\begin{table}[h]
\begin{minipage}{.5\textwidth}
\caption{ \small Dependence of $q$ on $a^*$.}
\begin{center}
\begin{tabular}{c|c|c|c|c|c}
\hline
\multicolumn{6}{c}{$\mu=0.3$, $\epsilon=0.02$, $\lambda=0.1$, $c=1$}\\ \hline
$q$ &  0.025 & 0.03     & 0.04     & 0.05 & 0.06 \\ \hline
$a^*$     &17.82 & 13.42& 8.42 & 5.33 & 3.18\\ \hline
\end{tabular}
\medskip
\end{center}
\end{minipage}
\begin{minipage}{.5\textwidth}
\caption{ \small Dependence of $\mu$ on $a^*$.}
\begin{center}
\begin{tabular}{c|c|c|c|c|c|c}
\hline
\multicolumn{7}{c}{$q=0.05$, $\epsilon=0.02$, $\lambda=0.1$, $c=1$}\\ \hline
$\mu$ &0.25 &0.3 & 0.4& 0.5& 0.6 & 1.1    \\ \hline
$a^*$ &3.97 &5.33 & 5.92 &5.7 & 5.3&3.72\\ \hline
\end{tabular}
\medskip
\end{center}
\end{minipage}
\begin{minipage}{.5\textwidth}
\caption{ \small Dependence of $\lambda$ on $a^*$.}
\begin{center}
\begin{tabular}{c|c|c|c|c|c}
\hline
\multicolumn{6}{c}{$\mu=0.3$, $q=0.05$, $\epsilon=0.02$, $c=1$}\\ \hline
$\lambda$   &0.05 & 0.12 & 0.15 & 0.17 & 0.2      \\ \hline
$a^*$ &4.84 &5.03&4.08&3.1&1.07\\ \hline
\end{tabular}
\medskip
\end{center}
\end{minipage}

\end{table}

\subsection{Rational Premium}
In this subsection, we consider the rational premium with $p(x)=c+1/(1+x)$.
One can solve  equation (\ref{eq:h}) and find the function $W_q$. If we take $w\equiv 0$ then, to get optimality of the barrier strategy using Theorem \ref{suffcond2}, we will assume that $\epsilon\leq q+\lambda$.
Thus, in the absence of the penalty function, we can find the values of $a^*$ for different parameters. In Tables 4, 5 and 6 we give some results in the case of a rational premium.
\medskip
\begin{table}[h]
\begin{minipage}{.5\textwidth}
\caption{ \small Dependence of $q$ on $a^*$.}
\begin{center}
\begin{tabular}{c|c|c|c|c}
\hline
\multicolumn{5}{c}{$\mu=0.3$, $\lambda=0.1$, $c=1$}\\ \hline
$q$   &0.005 & 0.01 & 0.015 &0.02        \\ \hline
$a^*$  & 37.03 & 23.98 &17.16 &12.77\\ \hline
\end{tabular}
\medskip
\end{center}
\end{minipage}
\begin{minipage}{.5\textwidth}
\caption{ \small Dependence of $q$ on $a^*$.}
\begin{center}
\begin{tabular}{c|c|c|c|c}
\hline
\multicolumn{5}{c}{$q=0.01$, $\lambda=0.1$, $c=1$}\\ \hline
$\mu$ &0.15 & 0.2 & 0.25 &0.3    \\ \hline
$a^*$ &0 &23.98 & 22.39&20.05\\ \hline
\end{tabular}
\medskip
\end{center}
\end{minipage}
\begin{minipage}{.5\textwidth}
\caption{ \small Dependence of $\lambda$ on $a^*$.}
\begin{center}
\begin{tabular}{c|c|c|c|c|c}
\hline
\multicolumn{6}{c}{$q=0.01$, $\mu=0.3$, $c=1$}\\ \hline
$\lambda$   &0.05 & 0.12 & 0.15 & 0.2 &0.25    \\ \hline
$a^*$ &17.73 & 20.55 & 20.8 &19.16 &13.29\\ \hline
\end{tabular}
\medskip
\end{center}
\end{minipage}
\end{table}

Note that $a^*$ seems to have similar properties in both linear and rational premium examples.
\section{Proofs}\label{sec:proof}

\subsection{Proof of the Verification Theorem \ref{verthm}}

The proof is based on a
representation of $v$ as the pointwise minimum of a class of
``controlled'' supersolutions of the HJB equation.
We start with the observation that the value function satisfies a dynamic programming equation.
\begin{lemma}\label{dynamic}
After extending $v$ to the negative half-axis by $v(x)=w(x)$ for $x<0$, we have,
for any stopping time $\tau$,
$$v(x)=\sup_{\pi\in \Pi} \E_x\bigg[ e^{-q \tau\wedge T} v(X^\pi_{\tau\wedge T})+\int_0^{\tau\wedge T}e^{-q t} \,dL_t^\pi\bigg].
$$
\end{lemma}

This follows by a straightforward adaptation of  classical arguments
(see  e.g. \cite[pp.~276--277]{AM}).
We will prove that $v$ is a supersolution of the HJB equation.

\begin{lemma}\label{eq:Vpilemma}
The process
\begin{equation}\label{eq:Vpi}
V^\pi_t := e^{-q(t\wedge T)}v(X^\pi_{t\wedge T}) +
\int_0^{t\wedge T} e^{-qs}\, dL^\pi_s
\end{equation}
is a uniformly integrable (UI) supermartingale.
\end{lemma}
\proof
Fix arbitrary $\pi\in\Pi$, $x\geq 0$ and $s,t\geq 0$
with $s<t$.
The process $V^\pi_t$ is $\mathcal{F}_t$-measurable,
and is UI. Indeed, by Lemma \ref{dynamic} we have
\begin{equation*}
\E_x[V^\pi_t]\leq  \sup_{\pi\in\Pi}\E_x\bigg[ e^{-q(t\wedge T)}v(X^\pi_{t\wedge T}) +
\int_0^{t\wedge T} e^{-qs} dL^\pi_s\bigg]=v(x).
\end{equation*}
Now by integration by parts, the non-positivity of $w$
and the no exogeneous ruin assumption
\begin{equation}
v(x) \leq
\E_x\bigg[\int_0^\infty q
e^{-qs} r^x_s \,d s\bigg] \leq
x+ \int_0^\infty e^{-qt}p(r^x_t)\,dt\leq (A+1)x+B,\label{affinedomination}
\end{equation}
where the function $r^x_t$ given in (\ref{rtx}) satisfies (\ref{speed}).

Let $W^\pi_t$ be the following value process:
\begin{align}
& W_s^\pi := \operatornamewithlimits{ess\,sup}_{\tilde\pi\in\Pi_s}
J_s^{\tilde\pi}, \qquad J_s^{\tilde\pi} := \E\bigg[
\int_0^{T^{\tilde\pi}} e^{-qu}\,dL^{\tilde\pi}_u +
e^{-q T^{\tilde\pi}}w(X^{\tilde\pi}_{T^{\tilde\pi}})\bigg|\mathcal{F}_s\bigg], \label{eq:Wspi}\\
& \Pi_s: = \left\{\tilde\pi = (\pi,\overline{\pi}) = \{L^{\pi,\overline{\pi}}_u, u\geq 0\}:
\overline{\pi}\in\Pi\right\},\qquad
L^{\pi,\overline{\pi}}_u: =
\begin{cases}
L_u^\pi, & u\in[0,s[,\\
L_s^\pi + L_{u-s}^{\overline{\pi}}(X^\pi_s), & u\ge s,
\end{cases}\notag
\end{align}
where $L^{\overline{\pi}}(x)$ denotes the process of cumulative dividends of the strategy
$\overline{\pi}$ corresponding to the initial capital $x$.

The fact that $V^\pi$ is a supermartingale is a direct
consequence of the following $\p$-a.s. relations:
\begin{itemize}
\item[(a)] $V_s^\pi = W_s^\pi$, \quad (b) $W_s^\pi\ge \E[W^\pi_t|\mathcal{F}_s]$, where $W^\pi$ is the process defined in (\ref{eq:Wspi}).
\end{itemize}

Point (b) follows by classical arguments,
since the family $\{J_t^{\tilde\pi}, \tilde\pi\in\Pi_t\}$ of random variables
is  upwards directed; see Neveu \cite{Neveu} and Avram et al. \cite[Lem. 3.1(ii)]{APP2} for details.

To prove (a), note that on account of
the Markov property of $X^\pi$ it also follows that
conditional on $X^\pi_s$,
$\{X^{\tilde\pi}_u-X^{\tilde\pi}_s, u\ge s \}$
is independent of $\mathcal{F}_s$. As a consequence,
the following identity holds on the set $\{s < T^{\tilde\pi}\}$:
\begin{multline*}
\E\bigg[\int_0^{T^{\tilde\pi}} e^{-qu}\,dL^{\tilde\pi}_u +
e^{-q T^{\tilde\pi}}
w(X^{\tilde\pi}_{T^{\tilde\pi}})\bigg|\mathcal{F}_s\bigg]\\
\begin{aligned}
&=
e^{-qs}\E_{X^{\pi}_s}\bigg[
\int_0^{T^{\overline{\pi}}} e^{-qu}\,dL^{\overline{\pi}}_u+
e^{-qT^{\overline{\pi}}}w(X^{\overline{\pi}}_{T^{\overline{\pi}}})\bigg]
+
\int_0^s e^{-qu}\,dL^{\pi}_s\\
&= e^{-qs} v_{\overline{\pi}}(X^\pi_s) + \int_0^s e^{-qu}\,dL^{\pi}_u,
\end{aligned}
\end{multline*}
and then we have the following representation:
$$
J_s^{\tilde\pi} = e^{-q(s\wedge T)}
v_{\overline{\pi}}(X^\pi_{s\wedge T})
+ \int_0^{s\wedge T}e^{-qu}\,dL^{\pi}_u,
$$
which completes the proof on taking the essential supremum over the relevant family of strategies.
\qed

We prove that the value function $v$ is a solution of the HJB equation.
We will denote by $\mathcal{G}$ the family of functions $g$  for which
\begin{equation}\label{eq:ggmartI}
M^{g,T_I}
:=\{e^{-q(t\wedge T_I)}g(R_{t\wedge T_I}),\;
t\geq 0\}, \quad T_I := \inf\{t\ge0: R_t\notin I\},
\end{equation}
is a supermartinagle for any closed interval $I\subset [0,\infty [$,
and such that
\begin{equation}\label{diffineq}
\frac{g (x)-g(y)}{x-y}\geq 1\quad \text{ for all }x>y \geq 0, \qquad g(x)\geq w(x) \quad\text{ for } x<0
\end{equation} and
$g$ is utimately dominated by some linear function.

\begin{lemma}\label{vHJB}
We have $v\in\mathcal{G}$.
\end{lemma}
\proof
Taking a strategy of not paying any dividends, by Lemma \ref{eq:Vpilemma} we find that
the process (\ref{eq:ggmartI}) with $g=v$ is a supermartingale.
We will prove that
\begin{equation*}
v(x)-v(y)\geq x-y\qquad \text{for all } x>y\geq 0.
\end{equation*}

Let $x>y$. Denote by
$\pi^\epsilon(y)$ an $\epsilon$-optimal strategy for the case $X^\pi_0=y$.
Then we take the strategy of paying $x-y$ immediately and
subsequently following the strategy $\pi^\epsilon(y)$ (note that such a strategy is admissible), so that the following holds:
$$
v(x) \ge x- y + v_{\pi^\epsilon}(y) \ge v(y) - \epsilon + x- y.
$$
Since this inequality holds for any $\epsilon>0$, the stated lower
bound follows.
Linear domination of $v$ by some affine function follows from (\ref{affinedomination}).
\qed

We now give the dual representations of the value function on a closed interval $I$.
Assume that $\mathcal{H}_I$ is a family of functions $k$
for which
\begin{equation}\label{tildemart}\WT M_t^{k,\pi}:=e^{-q(t\wedge \tau^\pi_I)}k(X^\pi_{t\wedge \tau^\pi_I}) +
\int_0^{t\wedge \tau^\pi_I} e^{-qs} dL^\pi_s
\end{equation} is a UI supermartingale
for $\tau^\pi_I:=\inf\{t\geq 0: X^\pi_t\notin I\}$
and $$k(x)\geq v(x)\qquad \text{ for }x\notin I.$$
Then
\begin{equation}\label{repr1}
v(x) = \min_{k\in\mathcal{H}_I} k(x)\qquad\text{ for }x\in I.
\end{equation}
Indeed, let $\pi\in \Pi$, $k\in \mathcal{H}_I$ and $x\in I$.
Then the Optional Stopping Theorem applied to the UI Dynkin martingale yields
\begin{align*}
k(x) &\geq \lim_{t\to\infty}\E_x\bigg [ e^{-q(\tau_I^\pi\wedge
t)}k(X^\pi_{\tau_I^\pi\wedge t}) + \int_0^{\tau_I^\pi\wedge
t}e^{-q s}\,dL^\pi(s)  \bigg]\\
&\geq \E_x\bigg[e^{-q\tau_I^\pi}v(X^\pi_{\tau_I^\pi}) +
\int_0^{\tau_I^\pi}e^{-qs}\,dL^\pi(s)      \bigg],
\end{align*}
where the convention $\exp\{-\infty\}=0$ is used.\\
Taking the supremum over all $\pi\in\Pi$ shows that $k(x)\geq v(x)$.
Since $k\in\mathcal{H}_I$ was
arbitrary, it follows that
$$\inf_{k\in\mathcal{H}_I}k(x)\geq v(x).
$$
This inequality is in fact an equality since $v$ is a
member of $\mathcal{H}_I$ by Lemma~\ref{eq:Vpilemma}.
The value function $v$ admits a more important representation from which the Verification Theorem
\ref{verthm} follows.
\begin{proposition}\label{repr2}
We have
\begin{equation}\label{eq:rep-g}
v(x) = \min_{g\in\mathcal G}g(x).
\end{equation}
\end{proposition}
\proof
Since $v\in\mathcal{G}$ in view of Lemma \ref{vHJB}, by (\ref{repr1}) it suffices to prove that $\mathcal{G}\subset \mathcal{H}_{[0,\infty[}$.
The proof of this fact is  similar to the proof of the shifting lemma \cite[Lem. 5.5]{APP2}.
For completeness, we give  the main steps.
Fix arbitrary $g\in \mathcal{G}$, $\pi\in\Pi$ and $s,t\geq 0$ with $s< t$.
Note that $\WT M^{g,\pi}$ is adapted
and UI by the linear growth condition and arguments  in the proof of Lemma \ref{eq:Vpilemma} and by \cite[Sec. 8]{APP2}.
Furthermore, the following (in)equalities hold true:
\begin{equation*}
\E\big[\WT M_t^{g,\pi}\big|\mathcal{F}_{s\wedge T}\big] \stackrel{(a)}{=} \lim_{n\to\infty}
\E\big[\WT M^{g,\pi_n}_t\big|\mathcal{F}_{s\wedge T}\big] \stackrel{(b)}{\leq} \lim_{n\to\infty}
\WT M^{g,\pi_n}_{s\wedge T} \stackrel{(c)}{=}
\WT M_{s\wedge T}^{g,\pi}
\stackrel{(d)}{=}
\WT M^{g,\pi}_s,
\end{equation*}
where the sequence $(\pi_n)_{n\in \mathbb{N}}$ of strategies is defined
by $\pi_n=\{L^{\pi_n}_t, t\geq 0\}$ with $L_0^{\pi_n} = L_0^\pi$ and
\begin{align*}
L^{\pi_n}_u &:=
\begin{cases}
\sup\{L^\pi_{v}: v< u, v\in \mathbb{T}_n\}, & 0<u < T, \\
L^{\pi_n}_{T-}, & u\ge T,
\end{cases}\\
 \mathbb{T}_n &:=\left(\left\{t_k:=s+(t-s)\frac{k}{2^n}, k\in
\mathbb{Z}\right\}\cup\{0\}\right)\cap \mathbb{R}_+,
\end{align*}
where the above $T$ is calculated for the strategy $\pi$.
Since $s$ and $t$ are arbitrary,
it  follows that $\WT M^{g,\pi}$ is a supermartingale, which will complete the proof.

Points (a), (c) and (d) follow from the Monotone and Dominated Convergence Theorems.
To prove  (b), let $T_i:=T\wedge t_i$,
 denote $ \WT M^{g,\pi_n}=M$,
$ L^{\pi_n}=L$ and observe that
\begin{align*}
&M_t - M_s =  \sum_{i=1}^{2^n} Y_i + \sum_{i=1}^{2^n}{Z_i},\quad\text{with}\\
&Y_i := e^{-q T_i}g\left(X^{}_{T_i-}\right) - e^{-q
T_{i-1}}g(X^{}_{T_{i-1}}),\\
&Z_i:=e^{-q T_i}(g(X^{}_{T_i}) - g(X^{}_{T_i-} ) + \Delta L_{T_i})\mathbb{I}_{\{\Delta L_{T_i}>0\}}.
\end{align*}
The strong Markov property of $R$ and
the  definition of $X^\pi$ imply
\begin{align}
\E[Y_i|\mathcal{F}_{T_{i-1}}]  &=  e^{-q T_{i-1}}\E\big[e^{-q(T_i - T_{i-1})}g(X^{}_{T_{i-}}) - g(X^{}_{T_{i-1}}) \big|\mathcal{F}_{T_{i-1}}\big] \notag \\
&= e^{-q T_{i-1}}
\E_{X^{}_{T_{i-1}}}[e^{-q\tau_i}g(R^{}_{\tau_i}) -
g(R^{}_0)],  \label{eq:YEDOOB}
\end{align}
with $\tau_i := T_i\circ\theta_{T_{i-1}}$, where $\theta$ denotes the shift operator.
The right-hand side of ~\eqref{eq:YEDOOB}
is  non-positive because $g\in \mathcal{G}$.
Furthermore, it follows from (\ref{diffineq})
 that all the $Z_i$ are
non-positive. The tower property of conditional
expectation then yields
$$
\E[M_t- M_s\,|\,\mathcal{F}_s] \leq 0.
$$
This establishes inequality (b) and the proof is complete.
\qed

\noindent
{\it Proof of the Verification Theorem \ref{verthm}.}
Since $v_\pi$ is absolutely continuous and dominated by an affine function, $v_\pi$ is in the domain of the full generator of~$R$. This means that the
process
$$v_\pi(R_{t\wedge T_I})e^{-\int_0^{t\wedge T_I} \frac{\mathcal{A}v_\pi(R_s)}{v_\pi(R_s)}\;ds}
$$
is a martingale for any closed interval $I\in[0,\infty [$. By (\ref{HJB}) it follows that $v_\pi\in\mathcal{G}$, which completes the proof.
\qed

\subsection{Proof of Lemma \ref{diff}}
Take any $x\geq 0$. Then fix $a>0$ such that $x<a$.
From the definition of $\w$ given in (\ref{exit1}), conditioning on the first claim arrival time $\sigma_1$, we obtain
\begin{equation}\label{prawo}
W_q(x)=e^{-(\lambda+q) h}W_q(r_h^x)+\lambda\int_0^h\int_0^{r_t^x}W_q(r_t^x-z)\, dF(z)\,e^{-(\lambda+q) t}\,dt,
\end{equation}
for $h$ small enough, so that $r^x_h<a$.
As $h\downarrow 0$ we find that $W_q$ is right-continuous at $x$.
Moreover, rearranging terms in  ($\ref{prawo}$) leads to
\begin{equation*}
\frac{W_q(r_h^x)-W_q(x)}{r_h^x-x}=\frac{1-e^{-(\lambda +q) h}}{h}\frac{h}{r_h^x-x}W_q(r_h^x)-\frac{h}{r_h^x-x}\frac{\lambda}{h}\int_0^h\int_0^{r_t^x}W_q(r_t^x-z)dF(z) e^{-(\lambda +q) t}\,dt.
\end{equation*}
Letting $h\downarrow 0$ we conclude that $W_q$ is  right-differentiable with derivative
\begin{equation}\label{prawa}
W_{q,+}^\prime(x)=\frac{1}{p(x)}\left((\lambda +q)W_q(x)-\lambda\int_0^{x}W_q(x-z)\,dF(z)\right).
\end{equation}
Now take any $x>0$. Equation ($\ref{prawo})$ can be rewritten as
\begin{equation*}
W_q(\tilde{r}_0^x)=e^{-(\lambda +q) h}W_q(x)+\lambda \int_0^h\int_0^{\tilde{r}_t^x}W_q(\tilde{r}_t^x-z)\,dF(z) e^{-(\lambda +q) t}\,dt,
\end{equation*}
where $\tilde{r}^x$ is a solution to the backward equation $d\tilde{r}_t^x=p(\tilde{r}_s^x)ds$, $ \tilde{r}_h^x=x$. We take $h$ small enough, so that $\tilde{r}_0^x\geq 0$.
We thus get left continuity and
\begin{equation*}
\frac{W_q(x)-W_q(\tilde{r}_0^x)}{x-\tilde{r}_0^x}=\frac{1-e^{-(\lambda +q) h}}{h}\frac{h}{x-\tilde{r}_0^x}W_q(x)-\frac{h}{x-\tilde{r}_0^x}\frac{\lambda }{h}\int_0^h\int_0^{\tilde{r}^x_t}W_q(\tilde{r}_t^x-z)dF(z) e^{-(\lambda +q) t}\,dt.
\end{equation*}
Letting $h\downarrow 0$ we see that $W_q$ is  left-differentiable  with derivative
\begin{equation}\label{lewa}
W_{q,-}^\prime(x)=\frac{1}{p(x)}\left((\lambda +q)W_q(x)-\lambda \int_0^{x-}W_q(x-z)\,dF(z)\right).
\end{equation}
Since $F$ is absolutely continuous,
($\ref{prawa}$) and ($\ref{lewa}$) imply that $W_q$ is continuously differentiable and satisfies~(\ref{eq:h}).
Using the same arguments and definition (\ref{exit2}) one can show that the function $\gqw$ is continuously differentiable and satisfies (\ref{eq:G}).
This completes the proof.
\qed

\subsection{On the Value Function for the Barrier Strategy}
Note that for the barrier strategy until the first hitting of the barrier $a$, the regulated process $X^{\pi_a}$ behaves like the process $R$. By the strong Markov property of the PDMP $R_t$ and by (\ref{exit1}) for $x\in[0,a]$ we have
$$
v_a(x)=
\frac{\w (x)}{\w (a)} v_a(a) + \E_x\big[e^{-q\tau^-_0}w(R_{\tau_0^-})\mathbb{I}_{\{\tau_0^-<\tau^+_a\}}\big].
$$
Moreover, again using the strong Markov property we can derive
$$\E_x\big[e^{-q\tau^-_0}w(R_{\tau_0^-})\mathbb{I}_{\{\tau_0^-<\tau^+_a\}}\big]=\gqw (x)-\gqw(a)\frac{\w (x)}{\w (a)}.$$
Hence
\begin{equation}\label{reprva}
v_a(x)=
\frac{\w (x)}{\w (a)}( v_a(a)-\gqw (a))+\gqw (x).
\end{equation}
We will prove that
\begin{equation}\label{vprime1}
v_a^\prime(a)=1,
\end{equation}
from which the assertion of Theorem \ref{barthm} immediately follows.

Note that for the barrier strategy $a$ we have
\begin{equation}\label{vax>a}
v_a(x)=x-a+v_a(a)\qquad\text{for }x>a.
\end{equation}
Take any $a>0$ and $x\in[0,a[$.
From the definition of $v_a$ given in (\ref{cost2}) and fixed $a$, conditioning on the first claim arrival time $\sigma_1$, we obtain
\begin{align}
v_a(x)&=e^{-(\lambda+q) h}v_a(r_h^x)+\lambda\int_0^h\int_0^{r_t^x}v_a(r_t^x-z)\, dF(z)\,e^{-(\lambda+q) t}\,dt \notag\\
&\quad+\lambda\int_0^h\int_{r_t^x}^\infty w(r_t^x-z)\, dF(z)\,e^{-(\lambda+q) t}\,dt,\label{prawova}
\end{align}
where $h$ is small enough (so that $r_h^x\in]0,a[$).
Letting $h\downarrow 0$ we find that $v_a$ is right-continuous at $x$ for all $x\in[0,a[$.
Moreover, rearranging terms in  ($\ref{prawova}$) leads to
\begin{align*}
\frac{v_a(r_h^x)-v_a(x)}{r_h^x-x}
&=\frac{1-e^{-(\lambda +q) h}}{h}\frac{h}{r_h^x-x}v_a(r_h^x)-\frac{h}{r_h^x-x}\frac{\lambda}{h}\int_0^h\int_0^{r_t^x}v_a(r_t^x-z)\,dF(z)\, e^{-(\lambda +q) t}\,dt \\
&\quad+\frac{h}{r_h^x-x}\frac{\lambda}{h}\int_0^h\int_{r_t^x}^\infty w(r_t^x-z)\,dF(z)\, e^{-(\lambda +q) t}\,dt.
\end{align*}
Letting $h\downarrow 0$ we conclude that $v_a$ is  right-differentiable on $[0,a[$ with derivative satisfying
\begin{equation}\label{prawava}
p(x)v_{a,+}^\prime(x)=(\lambda +q)v_a(x)-\lambda\int_0^{x}v_a(x-z)\,dF(z)-\lambda\int_{x}^\infty w(x-z)\,dF(z).
\end{equation}
Now take any $x\in]0,a]$. Equation ($\ref{prawova})$ can be rewritten as
\begin{align*}
v_a(\tilde{r}_0^x)&=e^{-(\lambda +q) h}v_a(x)+\lambda \int_0^h\int_0^{\tilde{r}_t^x}v_a(\tilde{r}_t^x-z)\,dF(z)\, e^{-(\lambda +q) t}\,dt \notag\\
&\quad+\lambda \int_0^h\int^\infty_{\tilde{r}_t^x}w(\tilde{r}_t^x-z)\,dF(z)\, e^{-(\lambda +q) t}\,dt,
\end{align*}
where $\tilde{r}^x$ is a solution to the backward equation $d\tilde{r}_t^x=p(\tilde{r}_s^x)ds$, $ \tilde{r}_h^x=x$. We take $h$ small enough, so that $\tilde{r}_0^x\geq 0$.
We thus get left continuity on $]0,a]$ and
\begin{align*}
\frac{v_a(x)-v_a(\tilde{r}_h^x)}{x-\tilde{r}_h^x}
&=\frac{1-e^{-(\lambda +q) h}}{h}\frac{h}{x-\tilde{r}_h^x}v_a(\tilde{r}_h^x)-\frac{h}{x-\tilde{r}_h^x}\frac{\lambda}{h}\int_0^h\int_0^{\tilde{r}_t^x}v_a(\tilde{r}_t^x-z)\,dF(z)\, e^{-(\lambda +q) t}\,dt\\
&\quad+\frac{h}{x-\tilde{r}_h^x}\frac{\lambda}{h}\int_0^h\int_{\tilde{r}_t^x}^\infty w(\tilde{r}_t^x-z)\,dF(z)\, e^{-(\lambda +q) t}\,dt.
\end{align*}
Letting $h\downarrow 0$ we infer that $v_a$ is  left-differentiable  on $]0,a]$ with derivative
\begin{equation}\label{lewava}
p(x)v_{a,-}^\prime(x)=(\lambda +q)v_a(x)-\lambda \int_0^{x-}v_a(x-z)\,dF(z)-\lambda\int_{x-}^\infty w(x-z)\,dF(z).
\end{equation}
Under the assumption that $F$ is absolutely continuous, the  function $v_a$ is differentiable on $]0,a[$.
Now we will prove that it is differentiable at $x=a$. Take $x=a$. Then from the definition of $v_a$, for $x=a$, conditioning on the first claim arrival time we obtain
\begin{align}
v_a(a)&=e^{-(q+\lambda)h}v_a(a)+e^{-\lambda h}\int_0^h e^{-qt}p(a)\,dt+\lambda\int_0^h\int_0^a v_a(a-z)\,dF(z)\,e^{-(q+\lambda)t}\,dt\notag\\
&\quad+\lambda\int_0^h\int_a^\infty w(a-z)\,dF(z)\,e^{-(q+\lambda)t}\,dt+\lambda p(a)\int_0^h\int_0^t e^{-qs} e^{-\lambda t}ds\,dt.
\label{eq:vaa}
\end{align}
Differentiating (\ref{eq:vaa}) with respect to $h$ and setting $h=0$ gives
\begin{equation}\label{eq:war}
0=-(\lambda+q)v_a(a)+\lambda\int_0^av(a-z)\,dF(z)+\lambda\int_a^\infty w(a-z)\,dF(z)+p(a).
\end{equation}
By setting $x= a$ in (\ref{lewava}) and using (\ref{eq:war}) we get  $v_{a,-}'(a)=1$.
This together with (\ref{vax>a}) proves that $v_a$ has a derivative at $a$ and (\ref{vprime1}) holds.

\subsection{Proofs of Necessary and Sufficient Conditions for Optimality of the Barrier Strategy }

\noindent
{\it Proof of Theorem \ref{ver2}.}
To prove sufficiency, we need to show that $v_{a^*}$ satisfies the conditions of the Verification Theorem \ref{verthm}.
From Theorem \ref{barthm} it follows that $v_{a^*}$ is ultimately linear.
Moreover, by the choice of the optimal barrier $a^*$ we know that $v_{a^*}^\prime (x)\geq 1$.
Finally,
by definition of  $W_q$ and $G_{q,w}$ and the strong Markov property of the risk process $R$ it follows that
$$
e^{-q(t\wedge T)}W_q(R_{t\wedge T\wedge \tau^+_{a^*}}),\qquad e^{-q(t\wedge T)}G_{q,w}(R_{t\wedge T}) $$ are martingales. Hence
$$e^{-q(t\wedge T)}v_{a^*}(R_{t\wedge T \wedge \tau^+_{a^*}})$$ is a martingale.
This means that $v_{a^*}$ is in the domain of the full generator of $R$ stopped on exiting $[0,a^*]$ and
that $(\cA -q)v_{a^*}(x)= 0$ for $x\leq a^*$.

To prove necessity we assume that condition (\ref{inq2}) is not satisfied. By the continuity of the function $x\mapsto (\cA -q)v_{a^*}(x)$ there exists an open and bounded interval $\mathrm{J}\subset]a^*,\infty[$ such that $(\cA -q)v_{a^*}(x)> 0$ for all $x\in \mathrm{J}$. Let $\tilde{\pi}$ be the strategy of paying nothing if the reserve process $X^{\tilde{\pi}}$ takes a value in $\mathrm{J}$, and following the strategy $\pi_{a^*}$ otherwise. If we extend $v_{a^*}$ to the negative half-axis by $v_{a^*}(x)=w(x)$ for $x<0$, we have
\begin{equation*}
 v_{\tilde{\pi}}(x) =
\begin{cases}
\E_x[e^{-qT_\mathrm{J}}v_{a^*}(R_{T_\mathrm{J}})], & x\in \mathrm{J}, \\
v_{a^*}(x), & x\not\in \mathrm{J},
\end{cases}
\end{equation*}
where $T_\mathrm{J}$ is defined by (\ref{eq:ggmartI}).

By the Optional Stopping Theorem applied to the process $e^{-qt}v_{a^*}(R_t)$, for all $x\in \mathrm{J}$, we obtain
\begin{equation*}
v_{\tilde{\pi}}(x)=\E_x[e^{-q T_\mathrm{J}}v_{a^*}(R_{ T_\mathrm{J}})]=v_{a^*}(x)+\E_x\bigg[\int_0^{ T_\mathrm{J}}(\cA -q)v_{a^*}(R_s)\,ds\bigg]> v_{a^*}(x).
\end{equation*}
This leads to a contradiction with the optimality of the strategy $\pi_{a^*}$ and the proof is complete.
\qed

\noindent
{\it Proof of Theorem \ref{suffcond}.} In the first step, we will show that
\begin{equation}\label{pom1}
\lim_{y\uparrow x} \,(\cA-q)(v_{a^*}-v_x)(y)\leq 0\qquad\text{for all }x>a^*.
\end{equation}
Let $x>a^*$. By the Dominated Convergence Theorem we obtain
\begin{align*}
 \lim_{y\uparrow x}\,(\cA-q)(v_{a^*}-v_x)(y)
&=p(x)(v'_{a^*}-v'_x)(x)-q(v_{a^*}-v_x)(x)\\
&\quad +\int_0^\infty\left[(v_{a^*}-v_x)(x-z)-(v_{a^*}-v_x)(x)\right]\lambda \,F(dz).
\end{align*}
By \eqref{va} we have:
\begin{itemize}
 \item[i.] $(v'_{a^*}-v'_x)(x)=0$.
\item[ii.]  $(v'_{a^*}-v'_x)(b)=\wprime (b) \left(\hp(a^*)-\hp(x)\right)\geq0$ for $b\in [0,a^*]$ by the definition of $a^*$.
\item[iii.]  $(v'_{a^*}-v'_x)(u)=\wprime(u)\left(\hp(u)-\hp(x)\right)\geq0$  for $u\in [a^*,x]$ by the assumption \eqref{as1}.
\item[iv.] $(v_{a^*}-v_x)(a^*)\geq0$, thus by iii, $(v_{a^*}-v_x)(x)\geq0$.
\item[v.] $(v_{a^*}-v_x)(x-z)\leq(v_{a^*}-v_x)(x)$ for all $z\geq0$ by ii and iii.
\end{itemize}
Thus we have shown \eqref{pom1}.

Now assume that \eqref{inq2} does not hold. Then there exists $x>a^*$ such that $(\cA -q)v_{a^*}(x)>0 $. By the continuity of $(\cA -q)v_{a^*}$ we deduce that $\lim_{y\uparrow x}(\cA-q)v_x(y)>0$, which contradicts \eqref{pom1}.
\qed

\noindent
{\it Proof of Theorem \ref{convex}.}
In view of Theorem \ref{ver2}, it follows that to prove optimality of $v_{a^*}$ we
need to verify that $g(x)\leq 0$ for $x>a^*$, where
\begin{equation}\label{g}
g(x):=\mathcal{A} v_{a^*}(x) - q v_{a^*}(x).
\end{equation}
Recall that
\begin{equation*}
g(x+a^*)=p(x+a^*)-qv_{a^*}(a^*)-qx +\lambda \int_0^\infty (v_{a^*}(x+a^*-y) - v_{a^*}(x+a^*))f(y)\,dy.
\end{equation*}
The desired assertion follows once the following three facts are verified:
(i) $g$ is concave on $\mathbb{R}_+\backslash\{0\}$, (ii) $g(a^*)=0$
and (iii) $g'(a^*)= 0$.

To show (i)
recall that $g(x)=0$ for all $x\leq a^*$ (see the proof of Prop. \ref{ver2}). Moreover,
denoting $k(x,y):=v_{a^*}(x+a^*-y) - v_{a^*}(x+a^*)$ and noting that
$\frac{\partial^2}{\partial x^2} k(x,y)=\frac{\partial^2}{\partial y^2} k(x,y)$,
we have
\begin{align*}
\frac{\partial^2}{\partial x^2} \int_0^\infty k(x,y) f(y)\,dy
&=  \int_0^\infty \frac{\partial^2}{\partial y^2} k(x,y) f(y)\,dy \\
&=
\frac{\partial}{\partial y} k(x,y)f(y)|_0^\infty - k(x,y)|_0^\infty
+\int_0^\infty k(x,y) f''(y)\,dy \leq 0
\end{align*}
since $v_{a^*}(x+a^*-y) - v_{x+a^*}(a^*)\leq 0$, $f''(y)\geq 0$, $\frac{\partial}{\partial y} k(x,0)=v_{a^*}' (x+a^*)=1$ and $f(0)=0$.

Point (ii) is straightforward,
and (iii) follows from the fact that $g'(x)=0$ for any $x< a^*$
and  $g$ is continuously differentiable.
\qed

\noindent
{\it Proof of Theorem \ref{suffcond2}.}
Let $g$ be defined by (\ref{g}).
Recall that by the definition of $a^*$ we have $g(a^*)=0$.
Moreover for $x\geq a^*$ we have
\begin{equation*}
g'(x)=p'(x)+\lambda\int_0^x v_{a^*}(y)f'(x-y)\,dy-\left(q+\lambda\right).
\end{equation*}
Note that in the case of $w\equiv 0$, $v_{x}\geq 0$ for all $x\geq 0$. Thus by assumption $g'(x)\leq 0$ for $x\geq 0$ and by Theorem \ref{ver2} the strategy $\pi_{a^*}$ is optimal.
\qed

\section{Conclusions}\label{sec:remarks}
In this paper, we solved the dividend problem with a penalty function at  ruin. We found some sufficient and necessary conditions for a barrier strategy to be optimal. Unfortunately, some of them, like (\ref{inq2}) and (\ref{as1}), may be difficult to verify. Moreover, we analyzed only single barrier strategies. Therefore one can consider ``multi-bands strategies'' (see \cite{APP2}). It would also be interesting to consider  the effect of adding fixed transaction costs that have to be paid when dividends are being paid. In the next step, it would be reasonable to examine the so called ``dual model'' with a negative premium function and positive jumps. In such a model the premiums are regarded as costs and claims are viewed as profits. Such a model might be appropriate for a company that specializes in inventions and discoveries (see \cite{avanzi}).  However, we leave these points for future research.

\begin{acknowledgements}
This work is partially supported by the National Science Centre under the grant  DEC-2013/09/B/ST1/01778.
The second author kindly acknowledges partial support by the project RARE -318984, a Marie Curie IRSES Fellowship within the 7th European
Community Framework Programme.
\end{acknowledgements}


\end{document}